\documentclass{aa}
\usepackage{graphics,epsf,epsfig,graphicx}

\begin{document}

\title{Unequal-mass galaxy merger remnants:\\ spiral-like morphology but elliptical-like kinematics}

\author{Fr\'ed\'eric Bournaud \inst{1,2}, Fran\c{c}oise Combes \inst{1} \& Chanda J. Jog \inst{3}}
 
\offprints{F. Bournaud \email{Frederic.Bournaud@obspm.fr}} 
\institute{
Observatoire de Paris, LERMA, 61 Av. de l'Observatoire, F-75014, Paris, France
\and
Ecole Normale Sup\'erieure, 45 rue d'Ulm, F-75005, Paris, France
\and
Department of Physics, Indian Institute of Science, Bangalore 560012, India}
\date{Received XX XX, 2004; accepted XX XX, 2004} 
\authorrunning{Bournaud et al.} 

\abstract{It is generally believed that major galaxy mergers with mass ratios in the range 1:1--3:1 result in remnants that have properties similar to elliptical galaxies, and minor mergers below 10:1 result in disturbed spiral galaxies. The intermediate range of mass ratios 4:1--10:1 has not been studied so far. Using N-body simulations, we show that such mergers can result in very peculiar systems, that have the morphology of a disk galaxy with an exponential profile, but whose kinematics is closer to that of elliptical systems. These objects are similar to those recently observed by Jog \& Chitre (2002). We present two cases with mass ratios 4.5:1 and 7:1, and show that the merging causes major heating and results in the appearance of elliptical-type kinematics, while surprisingly the initial spiral-like mass profile is conserved.
}

\maketitle


\section{Introduction}

Numerical simulations commonly show that the merging of two equal-mass spiral galaxies results in the formation of an elliptical galaxy with an $r^{1/4}$ radial profile consistent with observations (de Vaucouleurs 1953), and which is mainly supported by velocity dispersion rather than rotation (e.g., Barnes 1992). Galaxy mergers with mass ratios of 1:1 to 3:1 and 4:1 have been studied by Bendo \& Barnes (2000), Cretton et al. (2001) and Naab \& Burkert (2003): these lead to the formation of  boxy or disky elliptical galaxies, depending on the mass ratio. On the other hand, minor mergers with mass ratios of more than 10:1 have also been studied in numerical simulations. They correspond to the merging of a spiral galaxy with a dwarf companion, which has visible effects on the morphology and kinematics of the remnant: Walker et al. (1996) have found a significant heating and thickening of the disk. However, the remnant of a minor merger remains an exponential disk supported by rotation. 

These numerical works do not account for a recent observational result: Chitre \& Jog (2002) have analyzed a set of observed advanced merger remnants, and have classified them in two morphological classes. ''Class~I" remnants are elliptical-type galaxies with an $r^{1/4}$ radial profile. ''Class~II" systems have an exponential profile typical of spiral disks (Freeman 1970) ; a central bulge is present, too, but does not dominate the mass distribution, so that these systems cannot be simply regarded as elliptical galaxies with faint outer disks. Among the Class~II objects, Jog \& Chitre (2002) have pointed out systems with kinematical properties typical of elliptical galaxies: they have velocity dispersions as large or larger than rotation velocity, while spiral disks are usually supported by rotation. We will call these merger remnants with spiral-like morphology and elliptical-like kinematics ''hybrid" systems.

In this Letter, we test the hypothesis of Jog \& Chitre (2002), that the unexplored range of mass ratios of 4:1 to 10:1 could lead to the formation of these hybrid systems in galaxy mergers. We present N-body simulations of galactic encounters with mass ratios 4.5:1 and 7:1, and show that the remnants reproduce the mixed properties observed in the hybrid systems. Mergers with such unequal-mass ratios are very likely to occur, since low mass galaxies are much more numerous than large-mass ones (e.g., Binney \& Tremaine 1987).


\section{Numerical simulations}

\begin{figure*}
\centering
\resizebox{12cm}{!}{\includegraphics{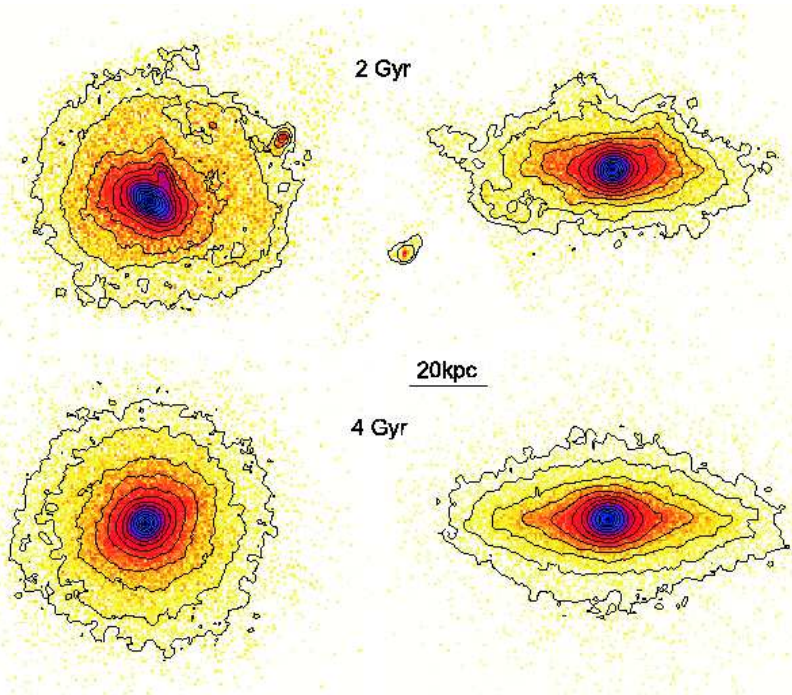}}
\caption{Snapshots of the 7:1 merger simulation (left: face-on --
right: edge-on), at different epochs. Top: 2 Gyr after the beginning of
the simulation, the system can be regarded as an advanced merger
remnant. The nuclei of the two galaxies have merged 400 Myr ago, but
strong asymmetries and tidal debris are still visible, as is the case
for the systems observed by Chitre \& Jog (2002). The morphological
and kinematical analysis (Figs.~2 and 3) has been done at this epoch, to
reproduce the observations of Chitre \& Jog (2002). Bottom: 2 Gyr later,
the system is fully relaxed. Its vertical light distribution resembles that of early-type spirals or S0 galaxies.}\label{maps}
\end{figure*}

We have used the N-body FFT code described in Bournaud \& Combes (2003). The gravitational potential is computed on a cartesian grid of size $256^3$ with a resolution of 700 pc. The number of particles is $10^6$ for the most massive galaxy, and is proportional to the mass for the other one. The dissipative gas dynamics is included using a sticky-particles scheme, described in detail in the Appendix of Bournaud \& Combes (2002), with parameters $\beta_\mathrm{r}=\beta_\mathrm{t}=0.8$. Star formation and time-dependent stellar mass-loss are modeled by the code described and the parameters used in Bournaud \& Combes (2002).

We have run two simulations with mass ratios 4.5:1 and 7:1. These ratios are that of the stellar masses. The mass of the main galaxy is $2\times 10^{11}$ M$_{\sun}$. Its stellar disk has a truncation radius of 15 kpc. It contains 8\% of gas, distributed in a disk of radius 30 kpc. The bulge and the dark halo are represented by two Plummer spheres of scale-lengths 3 kpc and 40 kpc respectively. The bulge-to-disk mass ratio is 0.3, and the halo-to-disk mass ratio inside the stellar radius is 0.5. For the companion, the radius of the stellar disk is scaled by the square root of its mass. It contains 11\% (4.5:1) or 13\% (7:1) of gas, distributed in a disk of radius 2.5 times the stellar radius. The bulge-to-disk mass ratio is 0.2, and the halo-to-disk mass ratio inside the stellar radius is 0.7. Disks are initially supported mainly by rotation, with small velocity dispersions corresponding to a Toomre parameter $Q=1.7$. 

The angle between the two disks, and the angle between each disk and the orbital plane, have been set to 33 degrees (which is the mean statistical value in spherical geometry). The impact parameter is set to 35 kpc, and the relative velocity $V$ of the colliding galaxies (computed for an infinite distance, assuming that dynamical friction is negligible before the beginning of the simulation) to 65 km s$^{-1}$. The orbit is prograde for the 7:1 merger, and retrograde for the 4.5:1 merger. The morphology and kinematics of merger remnants is then examined 400 Myr after the nuclei of two galaxies have merged at the resolution of the simulations, when the system corresponds to the advanced mergers studied by Chitre \& Jog (2002): single galaxies with strong asymmetries and tidal debris. The evolution of the 7:1 merger is shown in Fig.~\ref{maps}.

We have run a control simulation with the same main galaxy, that does not interact with any companion. This simulation is analyzed at the same epoch as the merger remnants, which enables us to disentangle the effects of the merger from other phenomena.

\section{Analysis of the simulated merger remnants}

\subsection{Morphology}

\begin{figure*}[]
\centering
\resizebox{5.6cm}{!}{\includegraphics{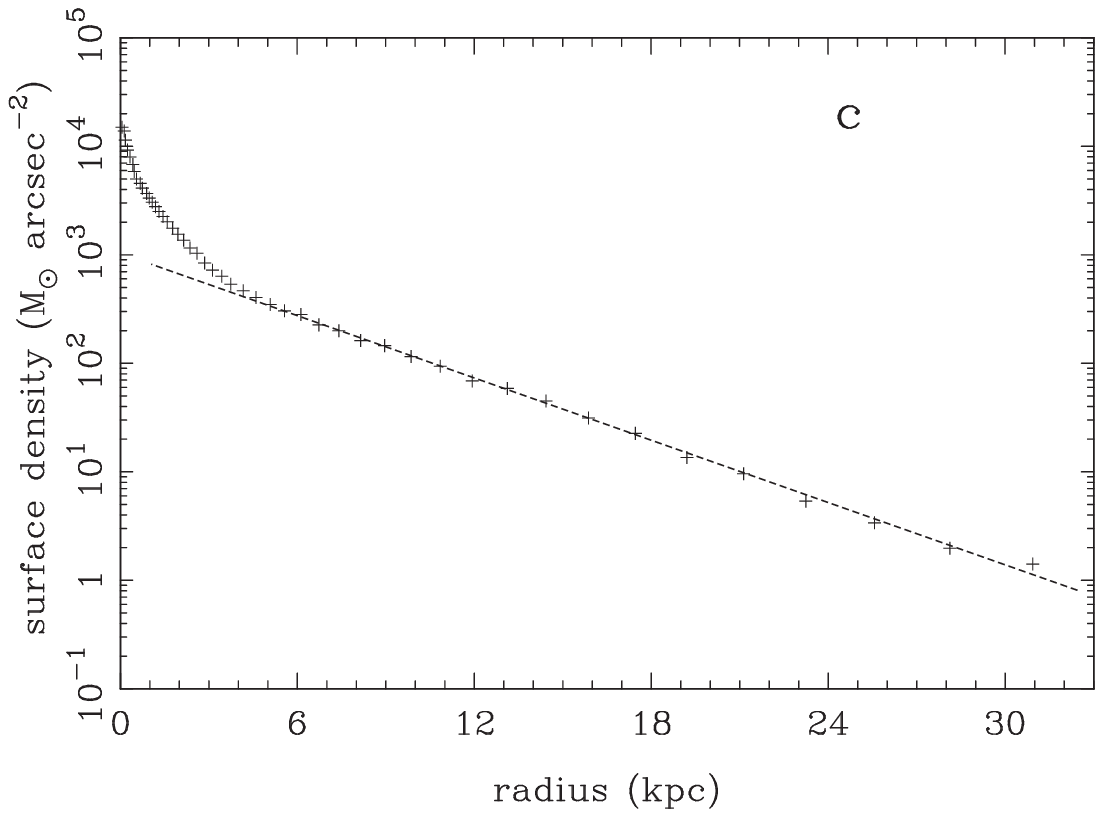}} 
\hspace{.4cm}
\resizebox{5.6cm}{!}{\includegraphics{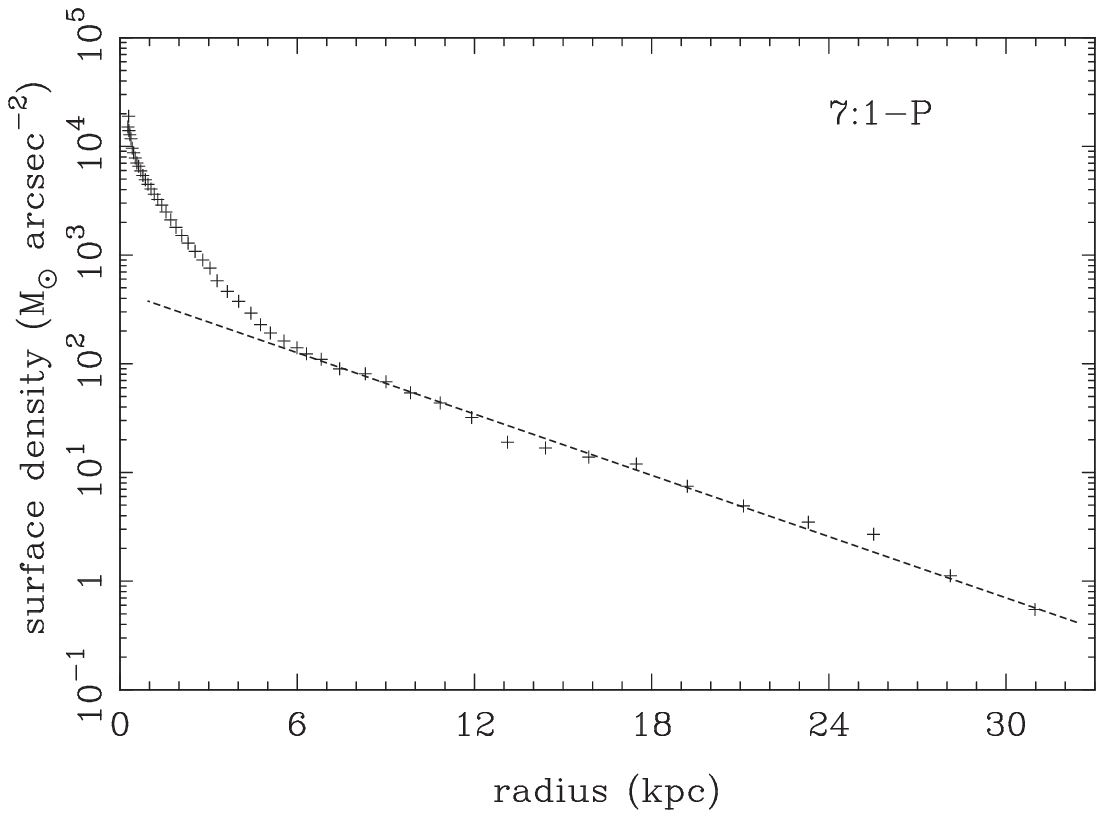}} 
\hspace{.4cm}
\resizebox{5.6cm}{!}{\includegraphics{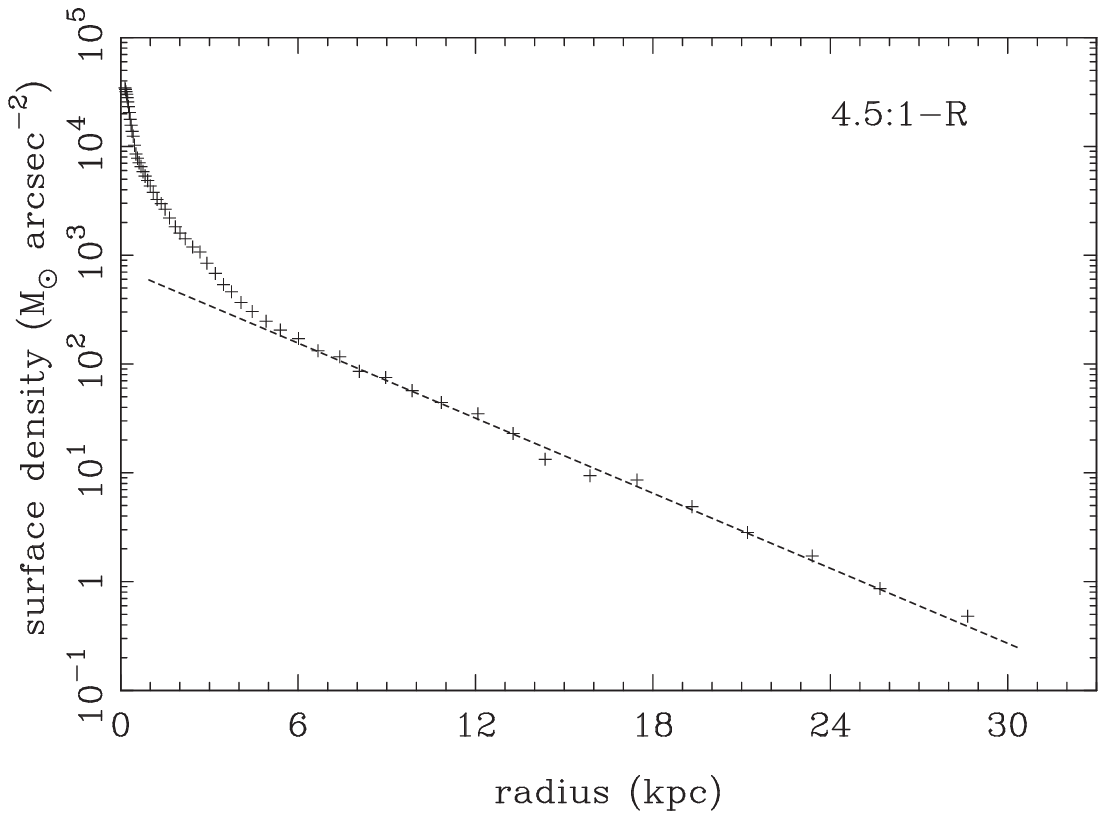}}
\caption{Radial density profiles of the stellar distribution for the control run (left panel) and the advanced merger remnants (7:1, prograde orbit: middle -- 4.5:1, retrograde orbit: right panel). The three systems show an exponential disk and a central bulge. We were unable to fit a robust $r^{1/4}$ profile on these data, while the e$^{-r/r_\mathrm{e}}$ profile is fitted over a range as large as 2.5--3 times the exponential scale-length $r_\mathrm{e}$, so that the exponential fit can be regarded as robust.}\label{profiles}
\end{figure*}

Based on the existing simulations of major and minor mergers (see Introduction), we expect the morphological transition between exponential disks and $r^{1/4}$ systems to occur in the new range of mass ratios that we have explored. The azimuthally averaged luminosity profiles of the stellar component in the 4.5:1 and 7:1 merger remnants are shown in Fig.~\ref{profiles}. They both show a massive exponential disk and a central bulge. The bulge is more massive by factor 1.5--2 in the merger remnants than in the control run, but the disk component always represents more then 60\% of the mass. Such radial light distributions are typical of early-type spiral galaxies. 

\subsection{Kinematics}

\begin{figure*}[]
\centering
\resizebox{5.6cm}{!}{\includegraphics{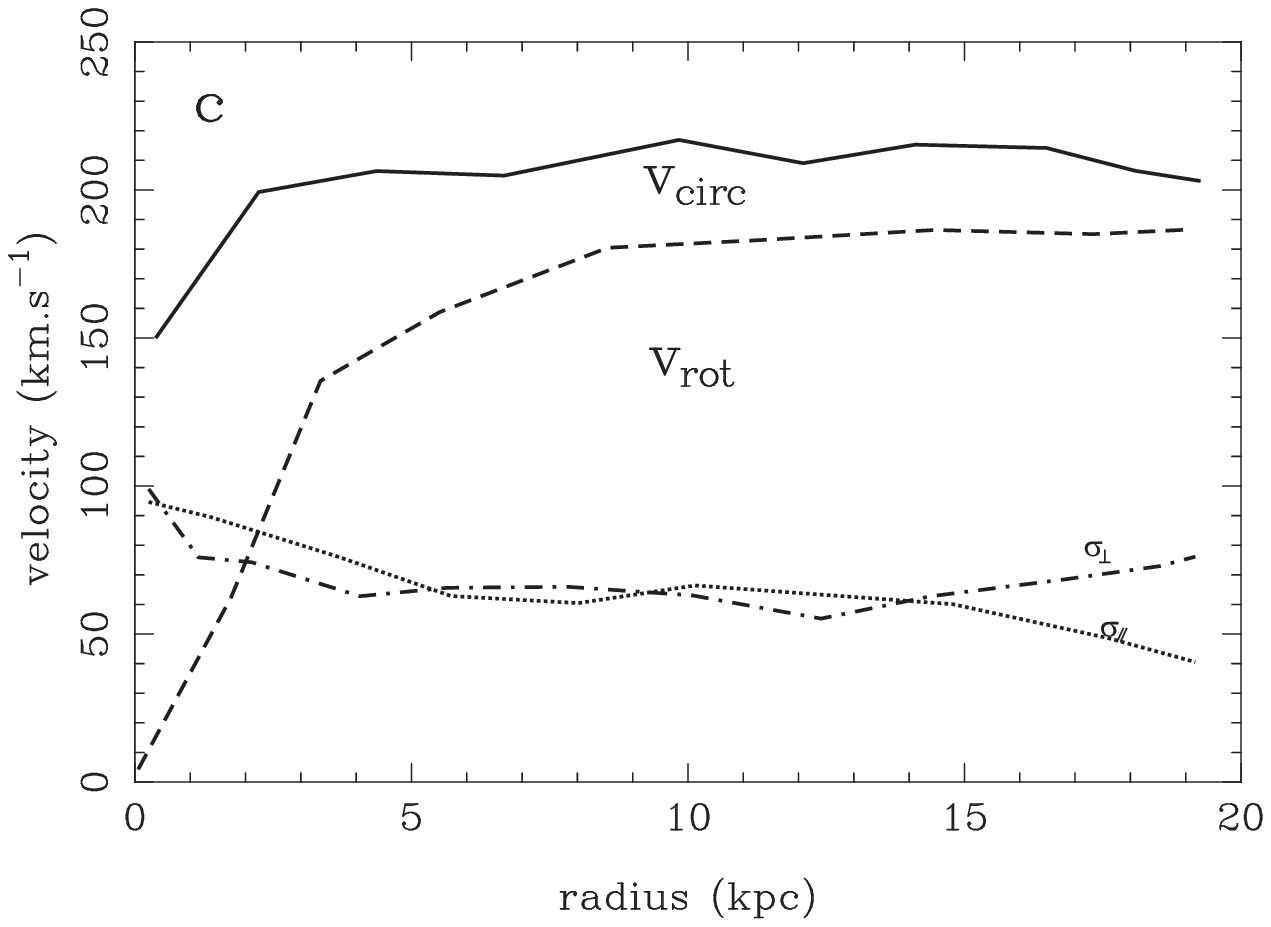}}
\hspace{.4cm}
\resizebox{5.6cm}{!}{\includegraphics{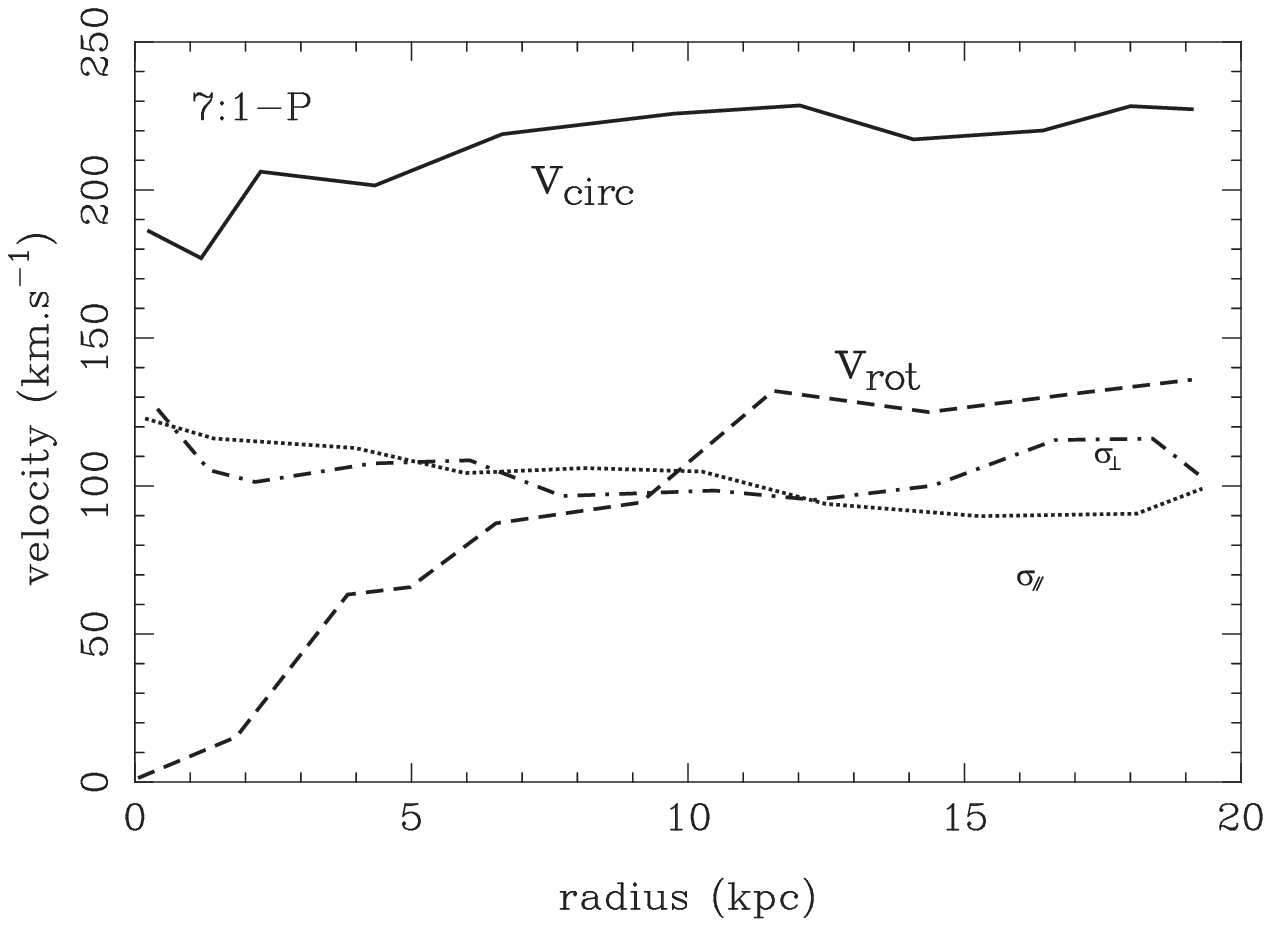}}
\hspace{.4cm}
\resizebox{5.6cm}{!}{\includegraphics{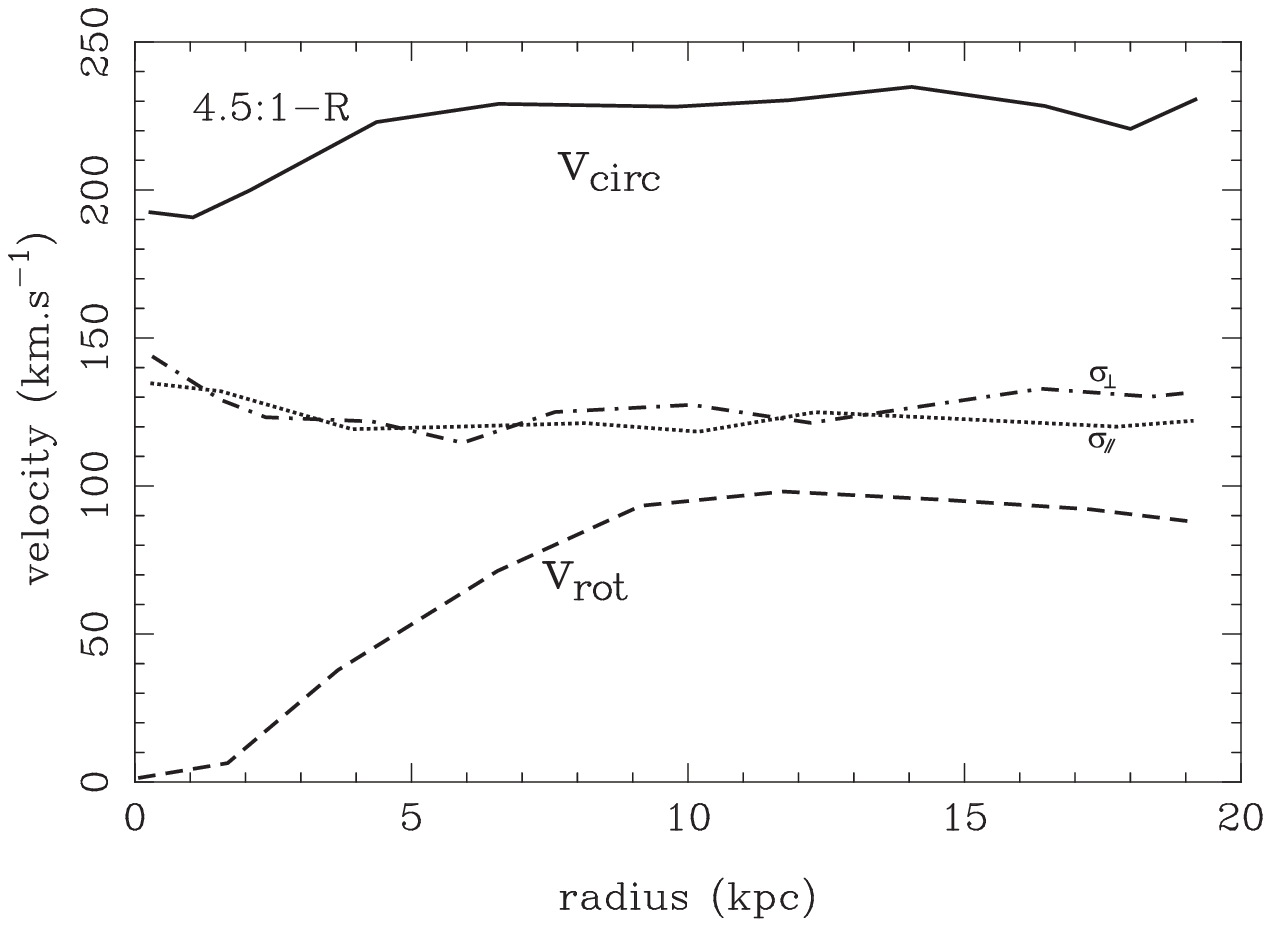}}
\caption{Kinematics of the control run (left panel), 7:1 remnant (middle), and the 4.5:1 remnant (right panel). The four curves represent the rotation velocity (dashed line), the velocity dispersions in the disk plane (dotted line) and perpendicular to it (dash-dotted line), and the circular velocity, i.e. the rotation velocity of stars that would have no dispersion (solid line). Outside the bulge component (4--6 kpc) and inside the isophotal radius (14--16 kpc), the value of $V_{\mathrm{rot}}/\sigma$ is typically 1 for the 7:1 case, and even smaller than 1 over the whole disk in the 4.5:1 merger remnant, which is characteristic of elliptical galaxies. The control run shows a much colder disk with $V_{\mathrm{rot}}/\sigma \simeq 3.3$. This proves that the appearance of elliptical-type kinematics cannot be regarded as an effect of secular evolution. 
}\label{kin}
\end{figure*}

The remnants of the 4.5:1 and 7:1 mergers show large velocity dispersions. For a mass ratio of 7:1, the value of $v/\sigma$ (rotation velocity to velocity dispersion ratio) is typically 1 for radii of 5 kpc (inner radius of the exponential disk component) to 16 kpc (isophotal radius), and smaller than 1.5 at all radii (see Fig.~\ref{kin}). For the 4.5:1 merger remnant, $v/\sigma$ is even smaller than 1 at all radii. The typical value in spiral disks supported by rotation is significantly larger than 1: in the control run we measure $v/\sigma \simeq 3.3$ even after the heating by bars and arms, and $v/\sigma$ can still be larger in cold disks (e.g., Binney \& Tremaine 1987). Velocity dispersions are even larger than rotation velocities over a 10 kpc radial range in 7:1 remnant, and over the whole disk in 4.5:1 remnant, that is mainly supported by velocity dispersions. This result is quite unusual, since at the same time these systems show a spiral-like morphology: usually, systems that are supported by velocity dispersions are elliptical galaxies with an $r^{1/4}$ profile, while exponential disks are kinematically much colder.

\subsection{Hybrid merger remnants}

The two galaxy mergers simulated here result in hybrid systems that have mixed properties, namely:
\begin{itemize}
\item a radial profile typical of spiral galaxies, with a massive exponential disk and a central bulge
\item kinematical properties of elliptical galaxies, with velocity dispersions of the same order or larger than the rotation velocities.
\end{itemize}
These mixed properties have been shown in Figs.~2 and 3 at the end of the merger, but are still observed a few Gyrs later (see Sect.~\ref{evo}). These reproduce well the observed hybrid merger remnants pointed out by Jog \& Chitre (2002). The unequal-mass merger scenario for the formation of such hybrid remnants, 
is thus proved to be robust.

\section{Discussion}
\subsection{Validity of the results}
A relevant question regarding the formation of hybrid systems in our simulations is whether the appearance of the elliptical-like kinematics is actually a consequence of the merger itself. Indeed, secular evolution with bars and spiral arms may significantly increase the velocity dispersions. However in the control simulation, the heating is very limited (see Fig.~\ref{kin}), in spite of the strongly barred spiral structure developed by the disk. In the outer disk, velocity dispersions only increase by 20--30\% with respect to the initial conditions. The heating is larger in the 5 central kpc, because of the bar, but dispersions still increase by a factor smaller than 2. The heating observed in the two merger remnants is 3 to 4 times larger for the same initial galaxy: it cannot be attributed to secular evolution. This also proves that the number of particles used and the softening of gravitational interactions used at short distances are properly adapted to avoid two-body scattering, that would have induced a major heating of the control run as well. Thus, the appearance of the elliptical-like kinematics can only be regarded as a consequence of the merger itself.

The second concern is the influence on the results of the schemes used for gas dynamics and star formation. The sticky-particles code that we have used has been discussed in Bournaud \& Combes (2002): we believe that it may over-estimate the loss of energy by the ISM only in the few central cells of the cartesian gird, i.e. only in the two or three central kpc, and we do not study these central regions here. Moreover, if ever the gas dynamics was not treated accurately, then this would result in the formation of new stars at the wrong place without changing the large-scale gravitational distribution of the old stars, but whether or not we account for the stars formed during and after the merger does not change our conclusions, since old stars largely dominate the total IR luminosity.
As for the star formation scheme, it may have a major importance in the central kiloparsec, where the gas infall induced by the merger initiates a large starburst. This central starburst, and the associated central luminosity peak, may largely depend on the modeling. Yet, this does not concern the disk component: outside the central kpc, the mass of stars formed during the merger is only a small fraction of the mass of pre-existing stars, so that our conclusions are not affected by the choice of the star formation parameters.

\subsection{Future evolution of the hybrid remnants}\label{evo}
We have also analyzed the remnants when they are fully relaxed, 2 Gyrs after the previous results. They still have a spiral-like radial distribution, and the same elliptical-like kinematics. When observed face-on (see Fig.~\ref{maps} for the 7:1 case), they show small distortions, especially lenses, but no strong bars or arms, as expected from their kinematics. Observed edge-on, they clearly show a disk component, with isophotes that are strongly disky. For the 4.5:1 merger remnant, we find $a_4=0.064\pm0.01$ over the exponential disk component, and $a_4\simeq0.07$ for the 7:1 merger (see e.g. Cretton et al. (2001) for the definition of $a_4$). In the control run, we measure $a_4\simeq0.09$, while the typical value of $a_4$ in disky elliptical galaxies is smaller, equal to $\simeq$ 0.01--0.02 (e.g. Naab \& Burkert 2003). Because of their large diskiness and their exponential profiles, these systems would be classified as disk galaxies. 

The hybrid merger remnants, once they are relaxed and appear isolated, have not evolved to elliptical galaxies in our simulations. In spite of their elliptical-type kinematics, they conserve a radial profile and a vertical structure typical of disk galaxies. However, their disk is 2 to 3 times thicker than in the control run, so that they finally could be described as ''thick and kinematically hot disk galaxies". Thus, these merger remnants could be regarded as candidates for S0 galaxies, for their bulge-to-disk mass ratio is typically 0.5, their velocity dispersions are larger by factor 2--3 than in spiral galaxies, and they are relatively gas poor: typically half of the gas is consumed by the central starburst, and a significant amount is removed in tidal tails, so that the hybrid remnants are much poorer in gas than the pre-existing spiral galaxies, even if not completely gas-depleted. Yet, a detailed comparison with observed S0s is beyond the scope of this Letter. With accretion of gas at the rate expected from the structure of spiral galaxies (Block et al. 2002), that would reform a thin, kinematically cold disk in a few Gyrs, such systems could also evolve to normal spiral galaxies embedded in thick and kinematically hot disks.


\section{Conclusion}
We have shown that galaxy mergers in the unexplored range of mass ratios 4:1--10:1 can result in the formation of hybrid systems, that have the mass distribution as in spiral galaxies and yet have kinematical properties closer to elliptical galaxies than spiral galaxies. They correspond to the merger remnants observed recently by Jog \& Chitre (2002). We have thus quantitatively proven their conjecture that for this mass range, the merger causes major heating and yet leaves the mass distribution exponential. 

Whether such hybrid systems are formed frequently should be studied with a large set of simulations, using various mass ratios and orbital parameters. This would more generally indicate the parameter range for which a galaxy merger results in an elliptical galaxy or a disk galaxy. This, and the evolution of hybrid remnants under successive galaxy mergers or external gas accretion, will be studied in future papers.


\begin{acknowledgements}
We are grateful to the anonymous referee whose comments have helped 
to improve the presentation of our results.
The simulations were computed on the Fujitsu NEC-SX5 of the CNRS computing center, at IDRIS. This work has been supported by the Indo-French grant IFCPAR/2704-1.
\end{acknowledgements}

\end{document}